%% file: main.tex
\documentclass{IEEEcsmag}

\usepackage[colorlinks,urlcolor=blue,linkcolor=blue,citecolor=blue]{hyperref}
\usepackage{comment}
\usepackage{tabu}                      
\usepackage{booktabs}                  
\usepackage{mwe}                       
\usepackage{xurl}
\usepackage{cite}
\usepackage{tikz}
\usepackage{rotating}
\usepackage{booktabs}
\usepackage{fontawesome}
\usepackage{pdflscape}
\usetikzlibrary{trees,positioning,shapes,shadows,arrows}
\usepackage{amsmath,amssymb,amsfonts}
\usepackage{tcolorbox} 
\usepackage{amssymb}
\usepackage{textcomp}
\usepackage{url}
\usepackage{adjustbox}
\expandafter\def\expandafter\UrlBreaks\expandafter{\UrlBreaks\do\/\do\*\do\-\do\~\do\'\do\"\do\-}
\usepackage{multirow}

\jvol{XX}
\jnum{XX}
\paper{8}
\jmonth{Jan/Feb}
\jname{Special Issue on Security and Privacy for the Metaverse}
\jtitle{Augmenting Security and Privacy in the Virtual Realm: An Analysis of Extended Reality Devices}
\pubyear{2024}

\newtcolorbox{boxH}{  
    boxrule = 0pt, 
    leftrule = 6pt 
}

\begin{document}

\sptitle{This is the author's version of the work. It is posted here for personal/educational use only. The definitive version was published in IEEE Security \& Privacy Magazine Jan/Feb 2024. [DOI:10.1109/MSEC.2023.3332004]}

\title{Augmenting Security and Privacy in the Virtual Realm: An Analysis of Extended Reality Devices}

\author{Derin Cayir}
\affil{Florida International University, Miami, FL, USA}

\author{Abbas Acar}
\affil{Florida International University, Miami, FL, USA}

\author{Riccardo Lazzeretti}
\affil{Sapienza University of Rome, Rome, Italy}

\author{Marco Angelini}
\affil{Sapienza University of Rome, Rome, Italy}

\author{Mauro Conti}
\affil{University of Padua, Padua, Italy}

\author{Selcuk Uluagac}
\affil{Florida International University, Miami, FL, USA}

\markboth{This is the author's version of the work. It is posted here for personal/educational use only.}{The definitive version was published in IEEE Security \& Privacy Magazine Jan/Feb 2024.}

\begin{abstract}\looseness-0 In this work, we present a device-centric analysis of security and privacy attacks and defenses on Extended Reality (XR) devices, highlighting the need for robust and privacy-aware security mechanisms. Based on our analysis, we present future research directions and propose design considerations to help ensure the security and privacy of XR devices. 
\end{abstract}

\maketitle

\input{1-Introduction}

\input{2-Methodology}

\input{3-Existing}

\input{4-SecurityAttacks}

\input{5-PrivacyAttacks}
\input{6-VE}
\input{7-Discussion}

\input{8-Conclusion}

\section{ACKNOWLEDGMENTS}
We thank the anonymous reviewers and special issue editors for their helpful feedback and time. This work was partially supported by (i) the US National Science Foundation (Awards: 2039606, 2219920), Cyber Florida, Microsoft, (ii) La Sapienza University of Rome within the Bando Ricerca 2020, project "Secure ANd PrivatE Information sharing (SAN-PEI)", and (iii) project "SERICS" (PE00000014) under the NRRP MUR program funded by the EU - NGEU. The views are of the authors only, not of the funding entities.

\def\refname{REFERENCES}

\vspace*{-8pt}

\begin{IEEEbiography}{Derin Cayir,}{\,} is a Ph.D. student at Florida International University (FIU), USA. Her research interests include privacy/security systems for XR devices. 
She is working as a Graduate Research Assistant in Cyber-Physical Systems Security Lab (CSL). She is an IEEE CS Student Member. Contact her at dcayi001@fiu.com.\vspace*{8pt}
\end{IEEEbiography}

\begin{IEEEbiography}{Abbas Acar,}{\,} is a Postdoctoral Associate at the CSL at FIU, USA. 
His research interests include privacy-aware technologies, alternative authentication methods, and security/privacy issues related to IoT. 
Contact him at aacar001@fiu.edu.\vspace*{8pt}
\end{IEEEbiography}

\begin{IEEEbiography}{Marco Angelini,}{\,} is an Assistant Professor in Engineering in Computer Science at Sapienza University of Rome, Italy. 
His research interests include Visual Analytics, applied in the Cybersecurity domain. Contact him at angelini@diag.uniroma1.it.\vspace*{8pt}
\end{IEEEbiography}

\begin{IEEEbiography}{Riccardo Lazzeretti,}{\,} is an Associate Professor at Sapienza University of Rome, Italy. 
His research focuses on security and privacy, with a particular focus on IoT. He is an IEEE senior member. 
Contact him at lazzeretti@diag.uniroma1.it. \vspace*{8pt}
\end{IEEEbiography}

\begin{IEEEbiography}{Mauro Conti,}{\,} is a Professor at the University of Padua, Italy, and he is also affiliated with TU Delft and the University of Washington, Seattle. His main research interest is in the area of Security and Privacy where he published more than 500 papers in topmost international peer-reviewed journals and conferences. 
He is a Fellow of the IEEE, AAIA, and Young Academy of Europe, and a Senior Member of the ACM. Contact him at mauro.conti@unipd.it.\vspace*{8pt}
\end{IEEEbiography}

\begin{IEEEbiography}{Selcuk Uluagac,} {\,} is currently an Eminent Scholar Chaired Professor in the Knight Foundation School of Computing and Information Science at FIU, where he leads the CSL. 
His research focuses on cybersecurity and privacy with 
practical and applied aspects where he holds hundreds of research publications in the most reputable venues. Contact him at suluagac@fiu.edu. 
\end{IEEEbiography}

\end{document}

%% file: 1-Introduction.tex
\chapteri{E}xtended Reality (XR) technologies stand at the forefront of a new digital revolution in an era of constant technological innovations. Nowadays, XR technology is much more than a device that produces 3D visuals. With new devices released each year and additional manufacturers getting involved in this field, the XR devices are considered for different application domains from entertainment to education to healthcare. The emerging metaverse realm offers a bright future with capabilities ranging from assisting astronauts in their mission to making hearing-impaired individuals "see" the conversations via subtitles.

XR devices are versatile in their functionality, equipped with an array of advanced sensors, communication capabilities, and hardware specifications. As these technologies evolve, our perception of reality seamlessly blends with the virtual world. However, the exponential growth of these technologies raises concerns about whether these devices are secure and the users' sensitive information is kept private. 
The increasing number of users will naturally attract attackers attempting to exploit these devices.
%
%
%
%
The challenge arises from the diverse sectors currently utilizing these technologies and the unique properties of the devices themselves.  
This heterogeneity of the devices 
aggregates the potential attacks, and complicates the examination of current devices.
Thus, it is vital for the research community in this field and the developers of these devices to consider what the current technologies propose and the vulnerabilities that the attackers can exploit.

%

\begin{figure*}[t]
\centering
  \includegraphics[width=.75\textwidth,clip, trim=0cm 0cm 0cm 0cm]{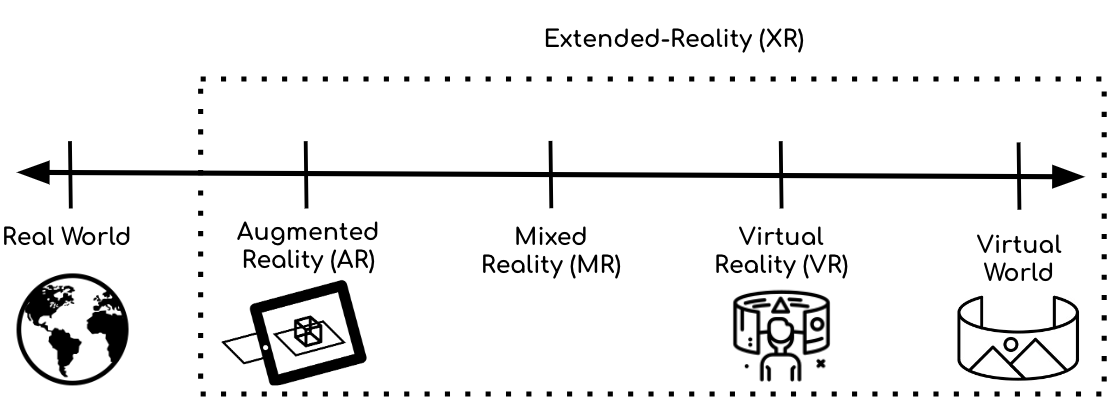}
  \caption{Spectrum of Extended Reality Technologies.}
  \label{fig:spectrum}
\end{figure*}
In this article, we study possible attacks on XR devices that could compromise the security and privacy of users and their environment in a device-centric approach. We highlight our key findings from detailed literature analysis, discuss the current attack vectors of XR devices, and present the security and privacy attacks with their corresponding defenses proposed in the literature. We analyze the attacks performed on the Virtual Environments (VE) separately, emphasizing the need for a further focus on this topic. Finally, we point to new research opportunities and propose design considerations, which can serve as valuable guidance for developers and the metaverse community. 

%% file: 2-Methodology.tex
\section{Methodology}

\subsection{Literature Review:}
To find the papers that perform security and privacy attacks or defenses on XR devices, we queried Google Scholar, ACM, and IEEE libraries on February 1, 2023. From 319 papers, we have restricted our selection to 41 papers listed in the table in Figure~\ref{fig:tablepapers}, testing practical attacks and their defenses that target XR devices' security and privacy. For interested readers, we detail our literature review methodology and the PRISMA 2020 guidelines followed further in the GitHub repository $^{3}$.
 
\subsection{Device Search:} 
From the selected papers, 
we gathered the devices used for the experiments. 
We also added to our device dataset other XR devices from the same companies that produced the devices mentioned in the papers. 
At the end of this process, in total, we identified 30 XR devices. The full list can also be found in the GitHub repository$^{3}$. 

\subsection{Security and Privacy Analysis:}
We examined the devices' security and privacy by analyzing their documentation websites, manufacturer posts, articles, and blogs, the links of which are given in the GitHub repository$^{3}$. In addition to the on-device properties, we analyzed the literature to find information about the security and privacy vulnerabilities of the devices and what types of attacks were seen on them. The questions we discuss in privacy policies are ``What type of data is collected, and where is it stored?'', ``Why is this data collected?'', ``Who is the data shared with?'', ``What are the users' rights on their data?'' and ``What are the privacy requirements of the apps on the devices?''.  

%% file: 3-Existing.tex
\section{Security and Privacy Mechanisms in XR Devices}
\label{section:Section3Mag}

\label{section:Section5}

In this section, we examine some general properties of XR devices. Then, we highlight XR devices' security and privacy mechanisms using their security documentation and privacy policies.

\subsection{General Properties of XR Devices}

Virtual Reality (VR) aims to replace the real world with a digital world, fully separating the user from their surroundings. On the other hand, Augmented Reality (AR) overlays virtual objects onto physical objects in the real world. Mixed Reality (MR) combines AR and VR, allowing interactive integration between the two worlds. XR encompasses AR, VR, and MR, containing all the devices that merge the virtual and real world, as shown in Figure \ref{fig:spectrum}. To seamlessly integrate the virtual world with the real world, XR devices strive to stimulate as many senses as possible (vision, hearing, smell, touch, and taste) through their sensors and actuators. Some general properties of XR that enhance realism are as follows:

\subsubsection{\textbf{Positional-tracking Features.}} 
XR devices offer 6 Degrees of Freedom (DoF) or 3 DoF, inside-out, or non-positional tracking. With 3 DoF the device can only track the rotational movement of the user, whereas with 6 DoF it tracks the user's rotation and position.
These are achieved by the inbuilt sensors such as gyroscope, magnetometer, accelerometer, cameras, infrared sensors, and IMUs.

\subsubsection{\textbf{Tracking Sensors.}} Tracking sensors play a crucial role in XR devices and they span from tracking the users' motions and interactions to their environment.

\noindent{\textit{User Motion Tracking:}} The XR devices have a Head-Mounted Display (HMD) that contains accelerometer, gyroscope, and magnetometer sensors to understand the head movements of the users. Similarly, the hand controllers are equipped with these motion sensors to track the position and orientation of the users' hands or even their finger movements. Many devices on the market, such as Meta Quest 2 and Microsoft HoloLens, support hand tracking where users can use their hands instead of a cursor. This is possible with inside-out cameras on the headsets$^{4}$. 

XR devices can also detect the users' body motions, tracking different body parts to translate these movements into avatars. For example, HTC sells Vive Trackers, external devices that users can attach to their bodies to integrate their movements into VR with more precise accuracy$^{5}$.

\noindent{\textit{User Interaction Tracking:}} Alongside translating body movements into the virtual realm, XR devices are equipped with eye and speech-tracking technologies which could be used to enhance the avatars, fully mimicking the users' speech and eye movements, and also developing more realistic simulations for medicine, missions, and much more. Microsoft HoloLens, Meta Quest Pro, and HTC Vive devices have eye-tracking sensors on the HMDs. With Vive Focus's eye tracker, users only need to gaze in a certain direction to open/close tabs or select objects$^{6}$. Meta Quest Pro also captures and stores the raw face-image of the users to extract the user's natural facial expressions to create more natural-looking avatars.$^{7}$

\noindent{\textit{Environmental Tracking:}} XR devices have outward-facing cameras that track everything within the users' environment, and facilitate the precise rendering of 3D objects into the user's environments. Proximity sensors detect the presence of objects, while depth sensors enable the devices to create a 3D map of the users' environment. Although VR devices are not primarily designed to integrate real-world and virtual-world objects as AR/MR devices are, many contemporary VR devices, including Meta Quest 2, Pico 4, PSVR 2, and Magic Leap, still incorporate these sensors and pass-through cameras to enable room-scale inside-out tracking. Meta apps can also use pass-through cameras to blend the physical and virtual environment of the users, a purpose that goes beyond merely viewing and not processing the real environment's data$^{8}$.

\subsubsection{\textbf{Audio and Speakers.}} 
Audio/speakers are integrated into the devices, and some devices have 3D spatial audio so users can physically locate the sounds they are experiencing in their virtual world. 
Meta Quest 2 and HTC Vive are examples of devices that use 3D spatial audio.

\subsubsection{\textbf{Haptic Feedback.}} 
Haptic feedback is an essential part of the VR experience to incorporate the users' senses into their virtual world. Different SDKs support haptics for developing immersive apps, such as vibrating the controllers$^{9}$ and applying force to simulate touch. There are also additionally sold suits, and gloves designed to make the metaverse experience even more realistic.

\subsubsection{\textbf{Communications.}} 
The XR devices include WiFi and Bluetooth communication so that users can collaborate with other users or connect to their other gadgets. Each device has its compatibility requirement and can run on different OS. For app development platforms, the devices are compatible with different graphic cards and RAMs.

\subsection{Security Properties of XR Devices}
\begin{figure}
\centering
  \includegraphics[width=\columnwidth,clip, trim=0cm 0cm 0cm 0cm]{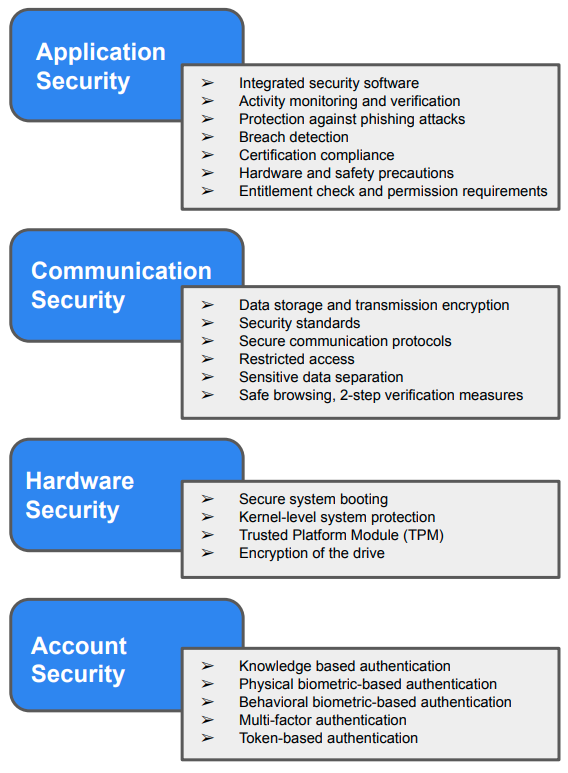}
  \caption{Security properties of XR devices.}
  \label{fig:secproperties}
\end{figure}

The impact of security and privacy attacks is high on XR devices as they are complex technologies that collect potentially user-identifiable information. Due to the immersive nature of these devices, attacks can manipulate users' perception of reality, potentially leading to physical harm. To ensure the security and privacy of the devices, vendors apply different methods that aim to meet the challenges of the modern cyber threats landscape, summarized in Figure \ref{fig:secproperties}. 

\noindent\textbf{Application Security.} Applications are essential for delivering different functionalities to users. Since applications have access to users' sensitive data, securing them against exploitation of sensitive information is a high priority. 
Device vendors adopt various measures to achieve this goal. Microsoft HoloLens relies on Microsoft Defender SmartScreen, integrated into the OS, warning users of dangerous websites and applications that can perform phishing and malware delivery. Meta monitors and verifies the account activity to prevent malicious acts and policy violations. 
Vuzix safeguards users' information from phishing attacks by preventing third-party apps from asking for user's sensitive information. 
Sony uses the information collected from the user to detect breaches such as unauthorized access to the apps. 
Pico Neo 4 uses ``ETSI EN 303 645''-based security certification,  which includes regular security updates and key management practices.

\noindent\textbf{Communication Security.}

Device vendors use different encryption standards to prevent attackers from accessing sensitive user information. For instance, 

HTC has data processing or altering, anonymization, pseudonymization, encryption during transmission using TLS, and access restriction. However, as stated in their policy, HTC does not take responsibility for threats from independent third-party applications. Microsoft HoloLens~2 secures data transfer between itself and the cloud using Azure integration. Furthermore, Dynamics 365 Remote Assist helps when deploying to external clients, separating sensitive device vendor data and resources. Google devices ensure continuous encryption to keep data private while in transit and have security features like Safe Browsing, Security Checkups, 2-step verification, physical security measures, and restricted access to personal info. Quest's VR messenger app prioritizes security, testing end-to-end encryption and specific apps' access control with its latest updates.

Similarly, Meta devices have end-to-end encryption. For digital audio and video content encryption, Epson Moverio supports HDCP-encrypted content. Additionally, the data in transit is protected through TLS by many other manufacturers, such as Magic Leap, Pico, Samsung, and Vuzix. 

\noindent\textbf{Hardware Security.}
Devices employ various hardware security measures to guard against unauthorized access and physical attacks. For instance, Microsoft HoloLens2 uses a Trusted Platform Module (TPM), a hardware-level security technology to generate, store cryptographic keys, and authenticate the device using unique RSA keys. Furthermore, BitLocker provides another level of security by encrypting the drive, employing AES-XTS-256 encryption, and safeguarding the data with multi-factor authentication, including Read Only (RO) media and privacy protection of writable data. 
Similarly, Pico devices use secure system booting, kernel-level system protection, and a trusted environment.

\noindent\textbf{Account Security.}
We note different authentication methods deployed on different devices. 
For instance, HTC Vive Cosmos and Epson Moverio use pin/password authentication. Quest Pro and Meta devices, on the other hand, use unlock patterns, also providing users with privacy customization options. Furthermore, some devices use accounts for login. Such as, Pimax links its authentication to a Steam account, Samsung authenticates through Oculus login, and Magic Leap relies on its ID system where a code verification is sent to a registered email address. 
Similarly, PSVR devices use a QR code for the initial device sign-in and then four-digit passwords with two-step verification for the remaining entries. Microsoft HoloLens supports iris-based authentication, but the users can also choose password entry to log in to their devices. Some third-party apps utilize biometrics, such as PalmID, which stores encrypted biometric signatures in Epson Moverio devices. On the other hand, the Pico Neo 3 Pro only requires login for the Pico App Store due to its business-focused purpose, where setup and access to files and apps must be quick.

\subsection{Privacy Policies of XR Devices}
\label{section:Section5.B}

The built-in sensors in XR devices collect data during or after the usage of the gadgets. Many of today's devices collect and share this information according to their privacy policies. So, we discuss the privacy of the devices in the current market by examining their policies and summarizing their properties in this section.

\noindent \textbf{What type of data is collected and how?}
Data collected by XR devices is highly sensitive, including information about users' physical properties, movements, environment, gender, age, gestures, and biometric information. If an attacker targets this data, the consequences can be damaging. 
Hence, users must be aware of the type of data the device vendors collect and where and how these data are stored. As stated in the Privacy Policies of Meta and HTC, the devices collect data in three ways: user-provided, sensor-collected, and third-party obtained. 
Information the users give while using devices may be about their transactions, social interactions, communications, email addresses, phone numbers, gender, location, physical features, avatar, content, and social media accounts.
Automatically collected data may be about people, games, apps, and features the users interact with. 
Through cookies, the data is linked to the user, including information about product access, device type, IP address, unique identifiers, WiFi network, web traffic, environment, physical dimensions (e.g., height, head size), play area, hand size, and movement. 
Information gathered from third parties may be from apps, developers, content providers, and marketing partners. 

\noindent\textbf{Where is the data stored?} Data storage practices vary between devices. Meta stores the data in the device in its raw form. 
Similarly, Magic Leap 2 has no cloud nor centralized server connection and stores the data on the device. 
HTC stores the data on the user's phone or HTC's servers (encrypting the data and not transmitting it anywhere other than the device and connected PC).

\noindent \textbf{Why is this data collected?}
Device vendors collect data for many purposes, including improving user experience, providing better-personalized services, communicating with the user, and protecting the manufacturers, its users, and the public (e.g., analyzing data to detect abuse, such as spam or illegal content). 
The data may also be used to enhance realism, such as using controllers, HMD movements, and audio to make the avatar more realistic.

\noindent \textbf{Who is the data shared with?}
Data collected by the devices can be shared without the users' knowledge. It is crucial for the users to understand what is done with their data and for developers of these devices to know how other vendors handle the data they collect. Generally, the data is shared with domain administrators, 
advertisement network providers, 
affiliated companies, 
other users, 
and third parties 
with the users' consent. 
Many vendors state that the data may be transferred, stored, and processed in any other country the device manufacturers business in less protective privacy laws. 

\noindent \textbf{What rights do users have over their data?}
Users have the right to manage, update, limit, and delete their data as well as to oppose and withdraw consent for data collection and marketing messages, as stated in Pico, HTC, and Vuzix's privacy policies. 
The user can do this by contacting the email provided on the website. 
Deleting a  Meta account results in deleting posts, entities, and apps, but not other users' posts about that user. 
With PlayStation VR (PSVR) devices, users can adjust the amount of shared data through the settings. 

\noindent \textbf{What are the privacy requirements of the apps on the devices?}
Most devices analyzed in this paper are programmable, where at-home users can create their own apps for their needs. However, this freedom comes with the cost of compromising the security and privacy of the devices. Developers should set basic app requirements to ensure a coherent experience and prevent malicious apps. Meta suggests Virtual Reality Check (VRC) guidelines for app developers in their Privacy Policy and requires the apps to follow their privacy policies, linking to the policy and clearly explaining collected data and use. 
Similarly, Google proposes general rules that app developers must follow for the users' safety. They define what can be collected from the users and how the apps should form their own privacy and content policies. 
Moverio prohibits collecting any information without users' consent and any phishing to gain sensitive information about the user. 

%% file: 4-SecurityAttacks.tex
\begin{figure*}
  \includegraphics[width=\textwidth,clip, trim=0cm 0cm 0cm 0cm]{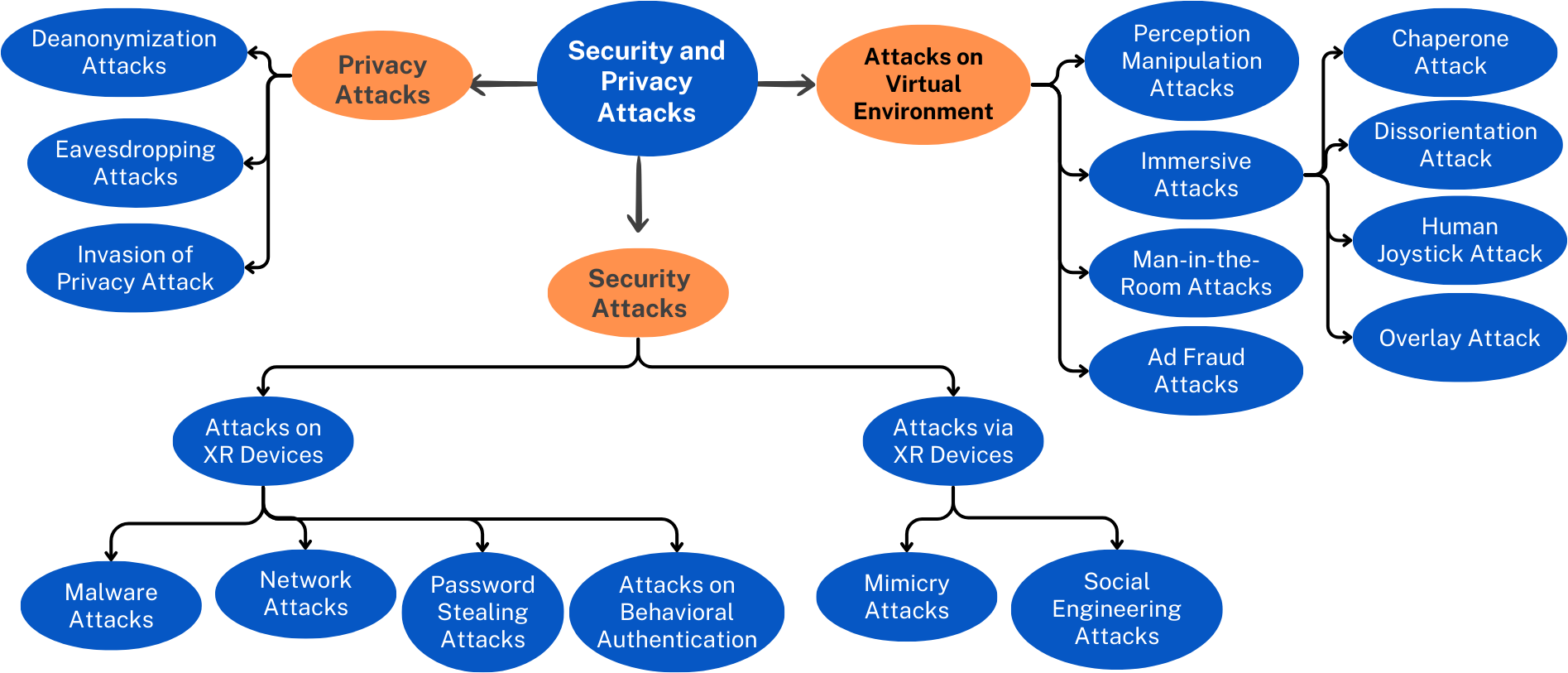}
  \caption{Security and privacy attacks in the literature.}
  \vspace{-10pt}
  \label{fig:attacks}
\end{figure*}

\begin{figure*}
  \includegraphics[width=\textwidth]{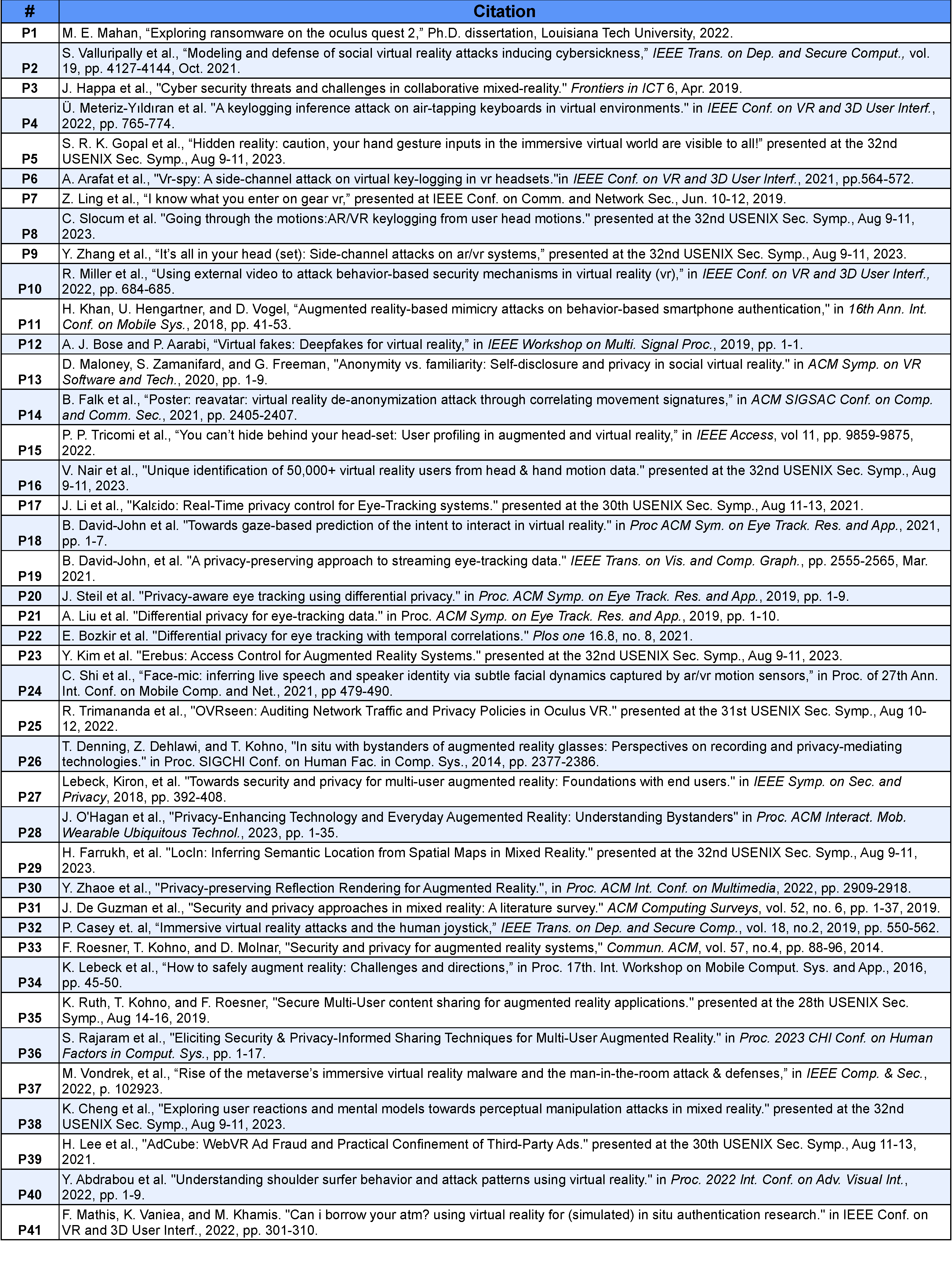}
  \caption{List of papers (additional references).}
  \vspace{-10pt}
  \label{fig:tablepapers}
\end{figure*}

\section{Security Attacks and Defenses}
\label{section:SecurityAttacksOverview}
 In this section, we categorize the security attacks into two categories: 1) Attacks on XR Devices and 2) attacks via XR devices. Our attack categorizations are shown in Figure~\ref{fig:attacks}. Papers presenting the attacks are listed in Figure~\ref{fig:tablepapers}. 
 
\subsection{Attacks on XR Devices}

\subsubsection{Malware Attacks} In this, an attacker plants viruses, or worms on users' devices without their knowledge. 
An example of a malware attack observed on VR devices is Big Brother, proposed by ReasonLabs.$^{10}$ 
The malware can infect VR devices with Android-based OS. 
With this, the attacker can remotely connect to an Android-based VR device and record the headset screen. This malware infects the user's computer, and once the malware enters the PC, it waits for a Developer-Mode-enabled VR device to connect. Upon connection, it opens a TCP port to record the user's headset whenever the PC and VR device share the same WiFi network.

Also, ransomware can target XR devices, limiting users' access until a ransom amount is paid$^{11}$. 
An Android ransomware sample was tested on Meta Quest 2 by integrating Simple Ransomware Sample (SRS) on the device, which is developed as a standard Android application [P1]. 
The goal was to get read and write data permissions through SRS, and encrypt the data with a function that uses Java Crypto and Security libraries. 
The researchers concluded that the attack surface of Meta Quest 2 includes essential elements that can be leveraged for effectively carrying out ransomware attacks. 

\subsubsection{Network Attacks}
In network attacks, the attackers exploit the vulnerabilities of the target network and bypass the security mechanisms in place. 
For instance, Valluripally et al.~[P2] showed a Denial of Service Attack (DoS) was executed via packet tampering, duplication, and dropping, resulting in the crashing of the virtual reality environment's server. 
Another study~[P3] also showed that 
DoS attacks resulting in the frame-rate drops in the devices may lead to nausea and dizziness, and create cybersickness attacks.

\subsubsection{Password Stealing Attacks}
Password-stealing attacks target the authentication of the devices and can lead to unauthorized access and 
sensitive information leakage. 
Key-logging attacks can be performed by capturing users' hand traces to identify their passwords while using an in-air tapping keyboard for input~[P4]. 
An adversary could plant hand tracker devices or videotape the users' text entry processes 
to obtain the victim's hand trace patterns and reconstruct inputs like passwords. This 
was also evident in a recent study where researchers retrieved graphical pattern lock inputs, passwords, emails, and pin entries of the users with VR HMDs, all from a video of the user interacting with the XR device~[P5]. 

Moving beyond visual observations, other studies explored non-visual approaches to identify key-logging~[P6]. These methods utilize a range of techniques, from analyzing network signals to leveraging device sensors, further highlighting the significance of this threat. 
For instance, Ling et al.~[P7] performed vision-based and motion-based side-channel attacks on Samsung Gear VR devices using sensors. 
The motion-based side-channel attack, in particular,  utilized the Samsung Gear VR's user motion tracking sensors by tricking the user into downloading a malicious app, 
which collects the orientation angles, hence giving information about where a key click occurs, and leading to the leakage of the user's password. Similarly, HMD's motion sensors from any XR devices with virtual keyboards revealed characteristics of users' typing behavior, enabling 
them to segment the motion signals and determine the typed words~[P8]. 
Moreover, a recent paper further exploited side-channel information such as thread times to differentiate digit inputs using a spy program on Microsoft HoloLens~2 and Meta Quest~2 devices~[P9].

\subsubsection{Attacks on Behavioral Authentication}
Due to usability considerations, behavioral authentication systems are considered ideal for AR/VR devices. Miller et al.~[P10] analyzed the ball-throwing task for authentication, where they could extract the 2D motion trajectories from the captured videos and match them to the 3D enrollment trajectories of users using HTC Vive, Vive Cosmos, and Meta Quest devices. 
Thus, they demonstrated that behavior-based authentication approaches could also be susceptible to attacks by obtaining the 2D video of the users. 

\subsection{Attacks via XR Devices}

\subsubsection{Mimicry Attacks}

AR devices can facilitate successful mimicry attacks on keystroke dynamics-like behavioral biometrics. Khan et al.~[P11] proposed an AR-based approach to mimic touch dynamics used for smartphone authentication. 
The results showed that 87\% of the attacks can bypass the authentication method. 

\subsubsection{Social Engineering Attacks}

The extensive data collection from advanced sensors of XR devices also poses security concerns through impersonation and social engineering attacks such as deep fakes. 
With deep fakes, an attacker could trick people into believing they are someone else. 
For instance, it has been shown that by creating deep fakes where the face and body of the user are physically altered to a digital form, the attackers can mimic the user and how they appear~[P12]. 
To prevent deep-fake audios, [P13] suggests that developers can make a layer to modulate the voice input obtained from the HMD so that the user's personal information is not identifiable.

%% file: 5-PrivacyAttacks.tex
\section{Privacy Attacks and Defenses}

\label{section:SecurityAttacks}
Privacy attacks focus on violating the users' right to exploit their personal information. 
In this section, we review the privacy attacks and defenses on XR devices.




\subsection{De-anonymization Attacks}
XR devices have many sensors that help navigate the user's environment, seamlessly blending real and virtual worlds. Naturally, these sensors collect information on the user's surroundings and personal information. Such information can be highly private as devices may record unique user movement data, potentially compromising their anonymity. 
For instance, in a de-anonymization attack called ReAvatar~[P14], users are identified by their virtual avatar via correlating specifically recorded movements. 
Remarkably, the user's movement remains unique even when using multiple avatars so that attackers can also de-anonymize them across multiple avatars. 
Moreover, [P15] shows that AR (Microsoft HoloLens) and VR (HTC Vive Pro) platforms are vulnerable to de-anonymization attacks, by identifying the users from basic physical actions like walking and pointing. 
In a more recent study~[P16], HMD and controllers' motion sensor data revealed user behavior patterns, making potential attackers re-identify users across different sessions of popular games. 

Another type of highly sensitive data is biometric data. 
Many XR devices - PSVR2, Magic Leap~2, Pimax Vision~8k, Microsoft HoloLens~2, and HTC Vive Pro Eye -  widely adopt eye tracking technology 
for different purposes ranging from authentication to understanding users' interests for advertising. Given the rich information content eyes offer, this raises critical privacy concerns. 
For example, pupil size can be used to understand someone's interests, while eye movements can be analyzed to infer mental disorders, cognitive states, gender, and age~[P17]. 
Researchers also found that the natural gaze dynamics from eye-tracking sensors could be used to predict the users' interaction with the virtual objects~[P18], and also by attackers for user identification~[P19]. 

Several strategies can be employed to safeguard user privacy, including the use of differential privacy, which involves the addition of random noise to obfuscate individual data without undermining overall data utility~[P17], [P20-22]. 
Currently, independent content developers can directly access raw data collected by XR devices' sensors. Hence, to protect the users' privacy, the researchers proposed designing APIs that would 
add Gaussian noise to raw data while also implementing temporal and spatial downsampling~[P19]. Furthermore, to prevent over-privileged malicious apps from accessing the raw sensor data, Kim et al.~[P23] propose an access control scheme for AR which allows users to limit the access to sensor data.

\subsection{Eavesdropping Attacks}
An eavesdropping attack tested in HTC Vive Pro and Meta Quest devices is the Face-Mic approach that derives sensitive information by exploiting motion sensors~[P24]. 
Speech-associated facial movements, bone-borne vibrations, and airborne vibrations of the user are collected, permitting to determine personal information, such as the user's gender. This attack utilizes zero-permission sensors (i.e., motion sensors) and reveals the user's protected information without the user's consent.

\subsection{Invasion of Privacy Attacks}
Invasion of privacy attacks involve the unauthorized collection and use of personal information. 
For instance, researchers realized that side-channel information could also be used for concurrent app fingerprinting on Microsoft HoloLens 2 devices~[P9], 
identifying which app the user is currently using. Furthermore, several academic works found that the traffic flow from the XR devices revealed user-identifiable information, especially when the users were using social applications~[P25].

Moreover, with outward-facing always-on cameras, the user themselves can record their environment in every detail without any notice, compromising the bystanders' privacy. Especially, AR glasses could be harder to notice in public settings, 
 where bystanders might not expect to be recorded [P26]. 
Several papers conducted user studies to test bystander privacy experiences in crowded public spaces~[P26-28], showing users' concerns about bystander privacy violations and invasive applications on their devices. 
In [P9], researchers found that environmental events created additional rendering, which was identifiable from the performance counter analysis of the devices. 
From this, the researchers identified the existence of a bystander by analyzing the CPU frame rates of Microsoft HoloLens, and they also calculated the distance of the bystander from the device. 

AR and MR devices capture spatial maps of the users' environment to overlay virtual content in the users' surroundings
by depth sensors and always-on cameras, which introduce privacy concerns. Researchers found that 
this can reveal information about the location of the users~[P29]. 
With a tailored malicious app, the researchers extracted the 3D spatial map of the user's environment using Microsoft HoloLens and identified the user's indoor location from a model trained with 3D objects present in an environment. Another study~[P30] found that inputs captured by AR devices during object rendering can contain sensitive objects, which will be translated onto the reflective AR objects. This reflection-based privacy attack results in the user's physical environment information being recovered by the attacker.

The literature suggests implementing defenses such as an intermediate layer between the sensor interfaces and the apps like input sanitization~[P31]. 
This way, sensitive information can be protected by the input access control system. This can be achieved in two different ways:



\noindent{\textit{Negotiating Permission:}}
Developers can include an option where the bystanders have a right to opt out if they feel their privacy is compromised~[P31]. 
For instance, physical switches to block the cameras or push-pull notifications where the bystanders near an XR device receive an option not to get recorded can be implemented~[P26]. 

\noindent{\textit{Blurring:}}
Developers can add a protection layer where sensitive objects (e.g., faces, license plates) in the captured images can be blurred~[P26].

%% file: 6-VE.tex
\section{Attacks and Defenses in Virtual Environment (VE)}

\label{section:AttackVE}
With XR devices, security and privacy concerns are not limited to the physical world. This section discusses the security and privacy issues in the VE.

\subsection{Immersive Attacks}
Immersive attacks target the unique properties of VR devices and are categorized into chaperone, disorientation, human joystick, and overlay attacks. A paper~[P32] shows this is possible in  HTC Vive and Oculus Rift devices by simply modifying VE parameters in a JSON file. 

\subsubsection{Chaperone Attack}
In a Chaperone attack, the attacker modifies the virtual boundaries of the victim~[P32]. 
In situations where the user's confidence in the boundaries that are no longer valid is high, the attacker might do physical harm to the user by altering the boundaries. A proof of concept attack was performed, and HTC Vive and Oculus Rift devices were found to be vulnerable against all tested OpenVR and SteamVR applications [P32]. 
To perform the chaperone attack, 
the researchers obtained the artifacts, such as the location of the VR boundaries, system settings, and executable path location, by exploiting SteamVR's vulnerability of storing the data in plain text without any integrity checks. 

\subsubsection{Disorientation Attack} 
In the disorientation attack, 
the user's location and rotation were adjusted by making minor changes in the player's orientation through yaw and translation parameters [P32]. 
In cases where the users are immersed in virtual environments and subject to visual motion cues without physical motion, Visually Induced Motion Sicknesses (VIMS) are seen. 
This way, the player's orientation is controlled, forming a seasick sensation. Smaller fluctuations in the artifacts resulted in stronger seasick sensations. These attacks were performed through Steam, and the success of this attack was similar to the Chaperone Attack as the same artifacts were targeted.

\subsubsection{Human Joystick Attack}
Human Joystick attacks are designed to alter the direction or location of a user within the VE without their awareness [P32]. 
These attacks aim to manipulate the user's movement, potentially leading to physical harm, such as the user hitting an object. 
For instance, the virtual environment was shifted continuously to move the user to an attacker-defined location. 
%
To solve these attacks, some countermeasures are suggested: intrusion detection, where an attack is flagged if it detects any patterns different from the expected timing model, or securing timing information, where the modulation frequency of the optical signal is changed. 

\subsubsection{Overlay Attack}
Attackers can superimpose images (such as inappropriate or alarming content) onto the user's screen to potentially cause harm or distress or block the user's view. Loud songs can be played, and bright flashing lights can be displayed on the XR device, which may harm users physically. 
These attacks can be particularly dangerous because users may not realize that the overlaid content is not part of the XR experience and may react to it as if it were real. 
An unlimited number of images was overlayed on Oculus Rift and HTC Vive devices [P32]. 
Furthermore, it is also found that packet sniffing attacks can be used to capture the users' physical location parameters illegally to perform overlay attacks [P2]. 


Overlay attacks are also a valid concern for AR devices that are designed to overlay computer-generated visual, audio, and haptic signals onto the real world~[P33]. In immersive AR applications, users must trust the app, and if it is targeted by the attacker, the users can be deceived about the real world. 
As a possible solution, windowing the display regions is suggested where the OS gives the applications separate windows corresponding to the bounded regions of the display~[P34]. 
With this solution, the applications' outputs are isolated from one another. Furthermore, [P34] proposes managing the outputs of AR devices as fine-grained objects, 
 made of first-class OS primitives, which make the OS capable of controlling when and where objects are placed. This method yields better flexibility and output control than windowing the displays. 

Security risks in AR do not just come from the apps themselves, but also from users who might intentionally spam others with disturbing virtual objects, or manipulate their virtual objects without permission. As a possible defense, Ruth et al.~[P35] propose an app-level library or an OS interface tailored for AR multi-user application developers. They consider the users' expectations, who may have different expectations about how AR content should be shared. Their proposed framework sets security objectives for controlling other users' permission to access shared (outbound) content and managing the incoming (inbound) content and owned physical space. They introduce "ghost" objects where certain sensitive parts of the object are not shared with other AR users, and they suggest policies on physical space ownership in AR. Furthermore, Rajaram et al. [P36] pairs AR, Security and Privacy experts to find solutions to AR overlay attacks. This study highlights that virtual menus and proximity-based interactions were suggested for content sharing and access control techniques. 

\subsection{Man-in-the-Room Attacks}
Man-in-the-Room (MITR) attacks represent a specific threat targeting the VE  where users are known to share private information [P37]. 
These attacks often exploit the users' immersion within the VE, benefiting from their tendency to assume the same privacy norms that are valid in the real world also hold in the virtual world. For example, a private virtual room that users may use to communicate with each other may be targeted by an MITR attacker as users would feel secure in a virtual room and would not expect an outsider to join without their consent. However, via MITR attack, the attacker can exploit this perception and know everything happening inside a private VR room without the victim's knowledge or authorization [P37]. 

An example of MITR attacks was performed on the Bigscreen VR app on Steam, which is supported by HTC Vive, Oculus Rift, and Windows Mixed Reality devices~[P37]. 
The Bigscreen app is used for communication in a VR environment. The attackers found a loophole where they exploited app vulnerabilities that caused a self-replicating infection (worm) without the user installing anything malicious. With the MITR attack, attackers could eavesdrop in the virtual room without other users noticing them. 
The attackers could turn on the users' microphones to listen to their conversations and observe their actions.

\subsection{Perception Manipulation Attacks}
Since the XR devices are designed to be highly immersive, many concerns have been expressed about the impacts of attacks on XR devices on the users. In~[P38], the researchers created three attack scenarios targeting visual, auditory, and situational awareness perceptions. 

With the visual attacks, the researchers overlay an adversarial virtual object, observing 
that the participants were fooled into believing the overlayed content was real, and their reaction times were significantly slowed. 
Interestingly, 
after the presence of the attacks, the participants started becoming hesitant and getting triggered by non-adversarial content. 
Imagine a real-world scenario where a user uses an XR device to get real-time guidance when driving and an adversary overlays incorrect speed limits, and traffic signs. The user will be deceived by these overlays and have a reduced reaction time, which is a valid concern while driving. The issue is that these attacks' impact continues even after the attack is finished as the users will lose their trust in the device and become hesitant with each traffic sign encountered.

Auditory attacks were performed 
while the users were concentrating on memorizing a sequence of elements. 
The immersive nature of XR audio led the users to treat the audio cues as a real-world stimulus. Lastly, 
the researchers displayed notifications on a screen that is in the background and realized that participants were not quick to notice real-world instructions while using their XR devices, showing that users are more focused in the metaverse than the real world.

\subsection{Ad Fraud Attacks}
Web-based VEs can be targeted by adversaries to create ad frauds that generate unintended ad traffic involving ad impressions or clicks [P39]. In XR technologies, the 3D world is rendered on an HTML canvas document object model (DOM) to create immersive experiences for the users and help them interact with the web page they are browsing. There are currently no primitives to separate the execution of an ad-serving JS script, enabling researchers to launch different attacks. One of these attacks was called gaze and controller jacking attacks, where a fake gaze and controller cursor were created to make the users intentionally click on the malicious VR objects. Furthermore, with a blind spot tracking attack, the researchers exposed the limited visual awareness of the users during 360-degree views by placing malicious promotional objects in the blind spots of the users' views. Similarly, with the abuse of an auxiliary display attack, the researchers could block the user from seeing their immersive world by displaying ads. As a potential solution, the researchers propose AdCube, which sandboxes the ad-serving JS and suggests that ad entities should be given a confined area. 
The researchers also suggest publishers specify DOMs that interact with a confined third-party ad script and generate access control policies on write and read permissions for DOMs.

%% file: 7-Discussion.tex
\section{Future Research Directions}
In this section, we leverage the insights gleaned from our study. 

\noindent\textbf{Authentication is the leading defense method.}
The current literature proposes unique ideas for user authentication, ranging from behavioral methods such as throwing a virtual ball to biometrics that utilize almost every part of the human body. Although authentication methods are the basis for securing the device from outsiders, because none of the devices have adopted the proposed authentication methods, it is clear that 
authentication offers only a partial panacea for device security.  

\begin{boxH}
    \textbf{Future Research Direction \#1:} Authentication alone cannot guarantee complete security, and it is important to consider multiple layers of security to address all possible attack vectors. 
    Therefore, 
    researchers must 
    propose additional defense strategies that tackle a broader range of security threats and vulnerabilities. 
\end{boxH}

\noindent\textbf{XR devices as virtual testbeds.}
Alongside XR devices serving as tools for various security attacks, they can also be used to create realistic virtual testbeds. This idea is explored in academia by generating scenes in XR devices to understand attacker behaviors [P40] and test the proposed methods' usability [P41]. VR-generated test environments provide remarkable similarities to real-world scenarios while addressing the shortcomings of in-person studies, such as overcoming ethical and legal constraints. Given their inherent flexibility, VEs are easily modifiable, making them ideal for such testing and educational scenarios.

\begin{boxH}
    \textbf{Future Research Direction \#2:} 
    Professionals across diverse disciplines 
    can utilize the XR devices to generate realistic testbeds and evaluate their algorithms within a remarkably authentic, yet controlled, environment.
    Additionally, VEs can facilitate testing the usability and efficiency studies of the defense solutions on the users.
\end{boxH}

\noindent\textbf{Device diversity in security testing.} The researchers predominantly utilize the same devices to apply their findings. The most used products were HTC Vive and Meta Quest~2 due to their wide accessibility and general public use. While we cannot assert that other devices not mentioned in this article are fully secure, we recommend that the readers focus on OS characteristics or examine the root causes of the vulnerabilities when understanding whether a type of attack is also applicable to their devices. 

\begin{boxH}
    \textbf{Future Research Direction \#3:} Future research should conduct security assessments using several devices, beyond just the popular ones. 
    This way, more attack vectors can be uncovered, identifying new potential vulnerabilities in a rapidly developing field.
\end{boxH}

\section{Design Considerations}

This section presents practical guidelines from our study to help developers create safer, more secure XR devices.

\noindent\textbf{Protection of sensitive data.} The immersive experience the XR technologies provide is made possible through the advanced sensors they are equipped with. However, 
our findings highlight that XR devices pose security and privacy risks by collecting intrusive sensor data, which can also expand the attack surface for other devices.  

\begin{boxH}
\textbf{Design Consideration \#1:} The accessibility of raw sensor data within XR device app development environments has established a notable threat model. 
    Therefore, we recommend developers of commonly used app development platforms (e.g., Unity, Unreal Engine) incorporate a default setting that limits the accessibility of users' raw data to independent app developers. Such as implementing differential privacy measures would protect the user data without compromising the app's performance.
\end{boxH}

\noindent\textbf{Physical input methods.} Input methods for sensitive data (e.g., passwords, text messages, emails) are highly physical as the user points the hands 
to a predefined location on a virtual keyboard. This opens up XR devices to numerous attacks, wherein an attacker will potentially extract the users' key presses, or replicate the authentication method by observing their actions. 

\begin{boxH}
    \textbf{Design Consideration \#2:} To prevent attackers from inferring the users' inputs, developers should utilize non-physical input methods. Eye-tracking technologies could be used for users to enter their passwords where they will enter their keys by looking at a key for a predetermined amount of time. Additionally, developers might consider methods like shuffling the keys of the keyboards to avoid virtual keyboard password-stealing attacks. 
\end{boxH}

\noindent\textbf{Virtual Environment as a new attack vector.} Security and privacy issues such as MITR attacks or inferring user passwords through user motions are specific to targeting the VE of a user.$^{5}$ While using XR devices, a user must continuously trust the environment generated by the devices. Hence, when an attacker targets the VE, the user who is fully immersed will be drastically affected.  

\begin{boxH}
\textbf{Design Consideration \#3:} In the design phase of virtual environments of VR and MR devices, the developers and device manufacturers should incorporate user feedback mechanisms. Utilizing insights from user studies on VEs, such as the one conducted by Lebeck et al. [P27], can provide an understanding of the users' needs, and expectations from the VEs. Additionally, direct features like in-app surveys can be done to further enhance user security. 
\end{boxH}


\noindent\textbf{Vague privacy policies.}  Several vendors' privacy policies are not explicitly tailored to individual devices and fail to distinguish between the data collected when using an HMD and other scenarios. Moreover, in current privacy policies, there is no explicit identification of who among the partners, developers, domain administrators, or affiliated manufacturers the data is shared with.


 \begin{boxH}
     \textbf{Design Consideration \#4:} Sensitive data collection by XR devices 
     requires clear communication and transparency from developers and manufacturers to users. Therefore, manufacturers should make their privacy policies easily accessible and understandable, communicating transparently about data collection and management processes. Features like opt-out options and data collection indicators should be added.
 \end{boxH}

%% file: 8-Conclusion.tex
\label{section:Section10}
\chapteri{I}n this article, we focused on the emerging technology of XR, conducting a comprehensive analysis of the security and privacy mechanisms of the devices currently dominating the market. 
Specifically, we provided an evidence-based approach where we analyzed the literature for security/privacy attacks on XR devices. We have also highlighted the critical need to analyze attacks and defenses in the VE. 
Lastly, we provided the lessons learned, which discuss the topics that could be further explored as future research and suggest some design considerations for developers to improve the security and privacy of their applications. Overall, this paper aims to help researchers understand what is currently needed as future defense directions and take appropriate measures.  